\documentclass{optica-article}
\journal{opticajournal}
\articletype{Research Article}
\usepackage{lineno}

\begin{document}

\title{Demonstration of $\bf3.5\times10^{-13}$ laser frequency stability at 1000\,s using an iodine-filled hollow-core fiber photonic microcell}

\author{Pengzhuo Wang,\authormark{1} Jose Sanjuan,\authormark{1} Moritz Mehmet\authormark{1}, and Felipe Guzman\authormark{1,*}}

\address{\authormark{1}James C. Wyant College of Optical Sciences, The University of Arizona, 1630 E. University Blvd., Tucson, AZ 85721, USA}

\email{\authormark{*}felipeguzman@arizona.edu} 

\begin{abstract*}
We present a laser frequency stabilization system based on an iodine-filled hollow-core photonic microcell (PMC), which is a sealed version of a hollow-core photonic crystal fiber (HC-PCF). A 532 nm laser is locked to the a1 component of the R(56) 32-0 transition of molecular iodine in the fiber cell, and its frequency stability is compared to that of the same component in a free-space iodine cell.
Noise analysis reveals that the system is limited by parasitic beams that interfere with the beam of interest and degrade the error signal. We have identified and characterized three types of parasitic interference and designed suppression methods for each. After applying these suppression methods, the frequency stability improved by more than an order of magnitude. The system achieves fractional frequency stability of $3.5\times10^{-13}$ for integration times around 1000 s. To our knowledge, this represents the best frequency stability achieved using a gas-filled hollow-core photonic crystal fiber frequency reference.
\end{abstract*}

\section{Introduction} \label{intro}
Laser frequency stabilization is crucial for many high-precision metrology applications, such as gravitational-wave detection~\cite{Cahillane:21}, laser ranging~\cite{PhysRevLett.geodesy}, atomic clocks~\cite{RevModPhys.87.637}, and many other interferometric applications~\cite{PhysRevApplied.7.024027}. Molecular or atomic frequency standards that utilize iodine, acetylene, rubidium, and other absorptive vapors have been extensively investigated for their inherent long-term stability~\cite{Schuldt:17,Edwards:04,Lemke2022}. Pushing for better stability has always been a direction of development, where as low as $10^{-15}$ level of fractional frequency stability was achieved using iodine cells~\cite{Zhang2024,Schuldt:17,4126932,Sanjuan:21}. The developments in other areas, such as optomechanical inertial sensing~\cite{Guzman2014}, increases the demand for compact optical frequency reference system. Efforts have been made to miniaturize free-space iodine cells for space applications~\cite{PhysRevApplied.Doringshoff,dringshoff2017}. However, gas-filled hollow-core photonic crystal fibers (HC-PCFs) offer a more promising approach, enabling compact all-fiber-based frequency reference systems~\cite{Benabid2005-rg, Lurie:12} that are particularly suitable for space applications where size and weight are critical.

HC-PCF filled with absorptive gases can be integrated into a compact, robust frequency reference system. However, the frequency stability of gas-filled HC-PCFs is generally worse than that of free-space cells. Previous investigations have shown that the short-term performance is limited by the linewidth of spectral features due to time-of-flight and power broadening~\cite{Lurie:11}. This is because the beam diameter inside the HC-PCF is usually on the order of tens of micrometers, which is two orders of magnitude smaller than the beam diameters in free-space gas cells. Higher beam intensity and more frequent energy dissipation from gas molecules to the fiber wall limit the frequency stability that can be achieved by HC-PCF systems. Regarding long-term stability, investigations of locking point repeatability suggest that HC-PCF systems are limited by high-order modes~(HOMs)~\cite{Knabe:09}. To mitigate this issue, single-mode operation of HC-PCF has been studied~\cite{Uebel:16,Vidal:25}. However, residual HOMs still exist and limit the system stability. Additionally, cladding modes that survive propagation are found to be spatially coherent~\cite{Couny:07}, meaning that cladding modes also contribute to frequency instability. In this paper, we include the cladding modes as part of the HOMs since both contribute to frequency stability degradation in a similar fashion. Additional studies have revealed that laser intensity noise and environmental factors, including airflow and temperature fluctuations, contribute to the system instability~\cite{Manamanni:22}. 

Previous investigations of gas-filled HC-PCF systems have been done with two ends of the fiber piece held in vacuum chambers. The adjustment of gas pressure inside the HC-PCF is more accessible, but it is further away from the purpose of an all-fiber-based system compared to a standalone fiber piece. The fabrication of sealed gas-filled HC-PCF, the so-called photonic microcell (PMC), has been studied by GLOphotonics SAS~\cite{Billotte:21}. One end of the HC-PCF is pre-endcapped. The closing of the other end involves a collapsing process that limits the insertion loss to 4\,dB~\cite{Goicoechea:23}. Because of the collapsing process, mode matching between the PMC and the input beam is problematic and, consequently, more input power and careful alignment are required to achieve similar performance with gas-filled PMCs compared to HC-PCFs. Moreover, coupling to the HOMs in the PMCs is harder to avoid, such that strong couplings from vibration and temperature fluctuation to frequency stability are expected. The previous best frequency stability levels achieved are $2.3\times10^{-12}$ at 1\,s integration time for iodine-filled HC-PCF~\cite{Lurie:11}, and $1.7\times10^{-12}$ at 1\,s for acetylene-filled HC-PCF~\cite{Chen:24}. For gas-filled PMC systems, performances are usually one order of magnitude worse compared to similar conditions in HC-PCF systems~\cite{Wang:13,Light:15} as the sealing process of the HC-PCF introduces artifacts and reduces the coupling efficiency. Environmental couplings to the frequency stability of gas-filled HC-PCF have not yet been fully characterized. Improving the performance of the gas-filled PMC frequency reference system will enable a broader range of applications.

To this end, we designed and tested a laser frequency stabilization system around an iodine-filled PMC. Different from a common saturation absorption spectroscopy system, a phase-lock loop and two optical isolators are introduced to mitigate the additional noise exhibited by the PMC. The main noise source limiting performance is the parasitic interference effect, which is caused by unwanted beams interfering at the photodetector (PD). After the mitigation of the parasitic interference effect, the frequency stability is improved by more than an order of magnitude. The system achieves stability levels below $10^{-12}$ for integration times between 20\,s and 2000\,s, which goes as low as $3.5\times10^{-13}$ at integration times of 1000\,s.

The paper is organized as follows: Sec.~\ref{system} describes the designed optical system, demodulation scheme, and feedback loops in detail. In Sec.~\ref{noise}, we characterize three types of parasitic interference and discuss noise suppression methods for each of them. Experimental results after applying noise suppression methods are discussed in Sec.~\ref{result}. Conclusions and future prospects are given in Sec.~\ref{conclusion}.

\section{System layout} \label{system}
\begin{figure}[htbp]
\centering\includegraphics[width=\linewidth]{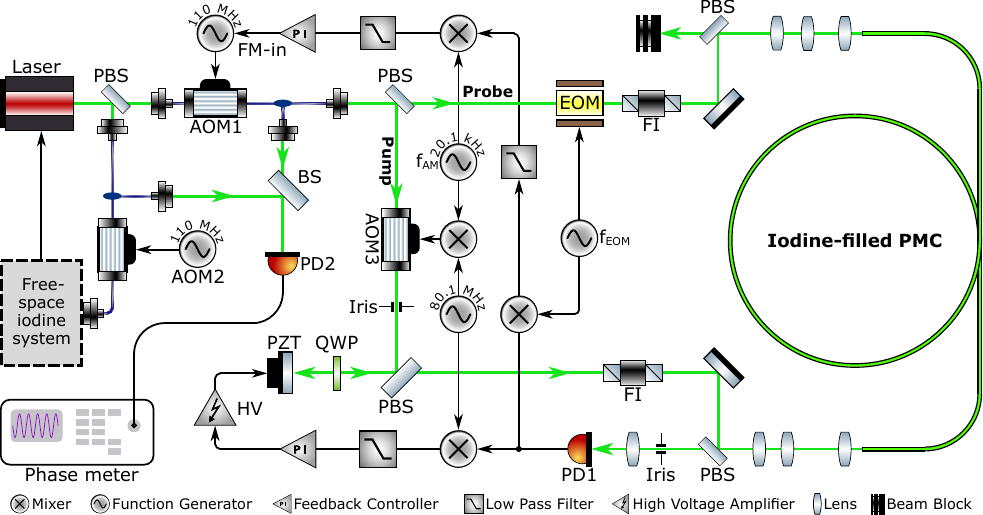}
\caption{Schematic diagram of the experimental setup (see text for details), including the optical system layout, laser modulation and demodulation, and feedback loops. PBS: polarizing beam splitter. BS: beam splitter. QWP: quarter-wave plate. AOM: acousto-optic modulator. EOM: electro-optical modulator. FI: Faraday isolator. PZT: piezo-electric transducer. PD: photodetector.}\label{fig: locking setup}
\end{figure}

Figure~\ref{fig: locking setup} shows the schematic diagram of the iodine-filled PMC system. A 532\,nm laser (Coherent, Prometheus 100) output is split into two. One path is guided into an iodine-filled PMC system, and the other path is used to lock the laser to a commercial free-space iodine system (TEM Messtechnik, FM Saturation Spectroscopy), which exhibits a fractional frequency stability below $10^{-13}$ for integration times longer than a few seconds~\cite{TEM_performance}. To characterize the stability of an iodine-filled PMC (GLO photonics, PMC-10-iodine-40-400), two acousto-optical modulators (AOM) are inserted before both setups: AOM1 (Brimrose, TEM-110-10-55-532-2FP) is used as the actuator to lock the iodine-filled PMC by frequency shifting the AOM driving signal. The frequency of the laser sent into the commercial unit is shifted by AOM2 such that the two systems can be locked to the same hyperfine structure of the iodine. A total optical power of 7 mW after AOM1 is sent to the PMC. A frequency-shifted (80.1 MHz) and amplitude-modulated (100\% modulation depth at $f_{\rm AM}$= 20.1 kHz) pump beam saturates the absorption of iodine gas. The frequency shift reduces the effect of pump leakage on the PD, and the amplitude modulation (AM) is applied to isolate the hyperfine structure of iodine, where the AM frequency is chosen to allow for sufficient controller bandwidth. The probe beam is phase-modulated~($f_{\rm EOM}=$ 60\,MHz, 1 rad modulation depth) by an electro-optical modulator (EOM, Qubig, PM8-VIS-60), and captures the dispersion of the Doppler-free structure of iodine. The signal measured by PD1 is first demodulated at $f_{\rm EOM}$ to obtain the differential dispersion of iodine. The second demodulation at the frequency of the amplitude modulation of the pump beam isolates the contribution of the hyperfines from the absorption background. The generated error signal is fed back to AOM1, thereby locking the laser frequency to the selected hyperfine component in the PMC. Beam splitters after AOM1 and before AOM2 are used to extract a fraction of the two beams stabilized via the PMC and the commercial free-space system, respectively. Both beams are brought to interfere on a 50:50 beam splitter and the beat note frequency is recorded with a commercial phasemeter (Moku: Pro, Liquid Instruments). Since the stability of the PMC system is expected to be one order of magnitude worse than that of the commercial free-space system, the beat note frequency is used to characterize the stability of the PMC system. 

\begin{figure}[htbp]
\centering\includegraphics[width=0.75\linewidth]{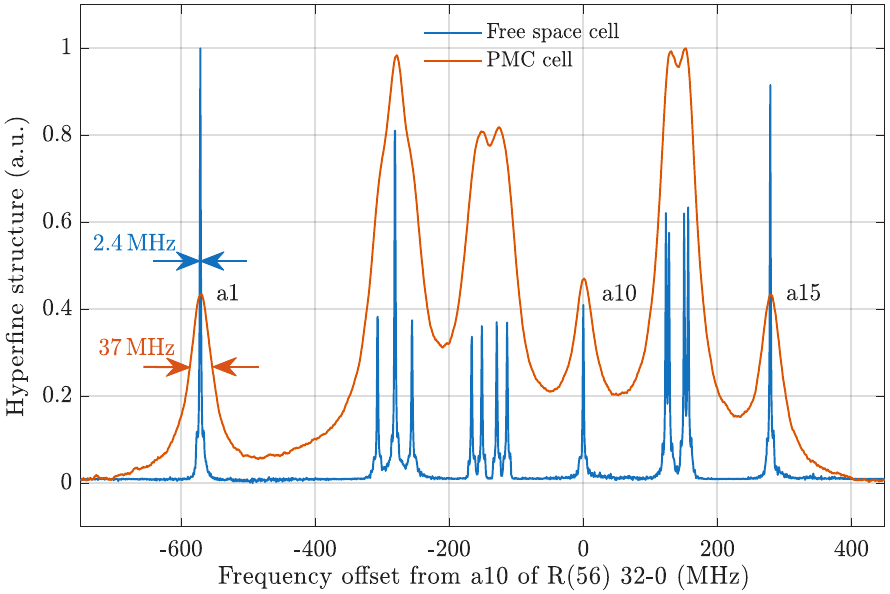}
\caption{Hyperfine structure of iodine R(56) 32-0 transition measured in the commercial free-space system and the PMC system. The linewidths of the a1 components are 2.4\,MHz and 37\,MHz in the free-space cell and PMC cell, respectively.}\label{fig: hyperfine}
\end{figure}

The discriminator (slope of the error signal at the zero-crossing) determines how electronic and shot noise map into frequency instability, and it is inversely proportional to the linewidth of the hyperfine structure. To estimate the stability that can be achieved, we measured the linewidth of the hyperfine structure in the iodine-filled PMC. The signal generated by PD1 is demodulated at $f_{\rm AM}$ while scanning the laser frequency. The resulting hyperfine structure of the iodine R(56) 32-0 transition is shown in Fig.~\ref{fig: hyperfine}. The linewidth of the a1 component of iodine in the PMC is 37\,MHz, which is an order of magnitude higher than the commercial free-space cell. The linewidth contribution from transit-time broadening is 9.5\,MHz based on the core size of the PMC. Other sources of the linewidth broadening are related to iodine-iodine collision~\cite{Light:15}, and power broadening due to the high intensity of the beams in the PMC. To prevent the phase-modulated beam from interacting with other hyperfine lines, we choose the a1 component as our target since it is 260 MHz away from the closest neighboring line.

\section{Noise sources and suppression methods} \label{noise}
Consider the generation of the error signal to lock the laser frequency to molecular iodine. The phase-modulated probe beam (field amplitude $A$, modulation depth $\beta$) can be separated into multiple frequency components, which consist of a carrier at nominal frequency ($\omega$) and sets of sidebands that are separated by the EOM modulation frequency ($\Omega_{\rm EOM}=2\pi f_{\rm EOM}$). For a small modulation depth, the electric field of the modulated beam is
\begin{equation} \label{eq1}
    E_{\rm EOM}=Ae^{i\omega t+i\beta\sin(\Omega_{\rm EOM} t)} \simeq Ae^{i\omega t}\left(1+\frac{\beta}{2}e^{i(\omega+\Omega_{\rm EOM})t}-\frac{\beta}{2}e^{i(\omega-\Omega_{\rm EOM})t}\right),
\end{equation}
where the carrier, upper, and lower sidebands experience different phase shifts ($\phi,\phi_+$, and $ \phi_-$) through the iodine-filled PMC, and interfere at the PD (gain and responsivity $G$). The voltage signal generated at the PD is
\begin{equation} \label{eq2}
    V_{\rm PD} = G|A|^2\left[\left(1+\frac{\beta^2}{2}\right)+\beta\cos(\Omega_{\rm EOM} t+\phi_+-\phi)-\beta\cos(\Omega_{\rm EOM} t+\phi-\phi_-)\right].
\end{equation}

By demodulating the signal at $\Omega_{\rm EOM}$, the differential dispersion of iodine is obtained. Assuming a unit amplitude of the local oscillator for demodulation, the resulting error signal near the center of the hyperfine line is
\begin{equation} \label{eq3}
    V = G|A|^2\frac{\beta}{2}[\sin(\phi_+-\phi)-\sin(\phi-\phi_-)].
\end{equation}

Due to the presence of parasitic beams at the same frequency as the probe beam carrier, additional uncertainty is presented in the error signal, i.e., 
\begin{equation} \label{eq4}
    V_\epsilon = G|A_\epsilon||A|\frac{\beta}{2}[\sin(\phi_+-\phi_\epsilon)-\sin(\phi_\epsilon-\phi_-)],
\end{equation}
where $A_\epsilon$ is the field amplitude of the parasitic beam at the PD that interferes with the phase-modulated beam (the probe beam). The contrast reduction due to the overlap of the polarization states and spatial modes is included in $A_\epsilon$. $\phi_\epsilon$ is the phase of the parasitic beam. For simplicity, we only include the first demodulation. In the actual system, the parasitic beam needs to be phase-modulated or amplitude-modulated at $f_{\rm AM}$ to survive the second demodulation described in Sec.~\ref{system}. 
\begin{figure}[htbp]
\centering\includegraphics[width=\linewidth]{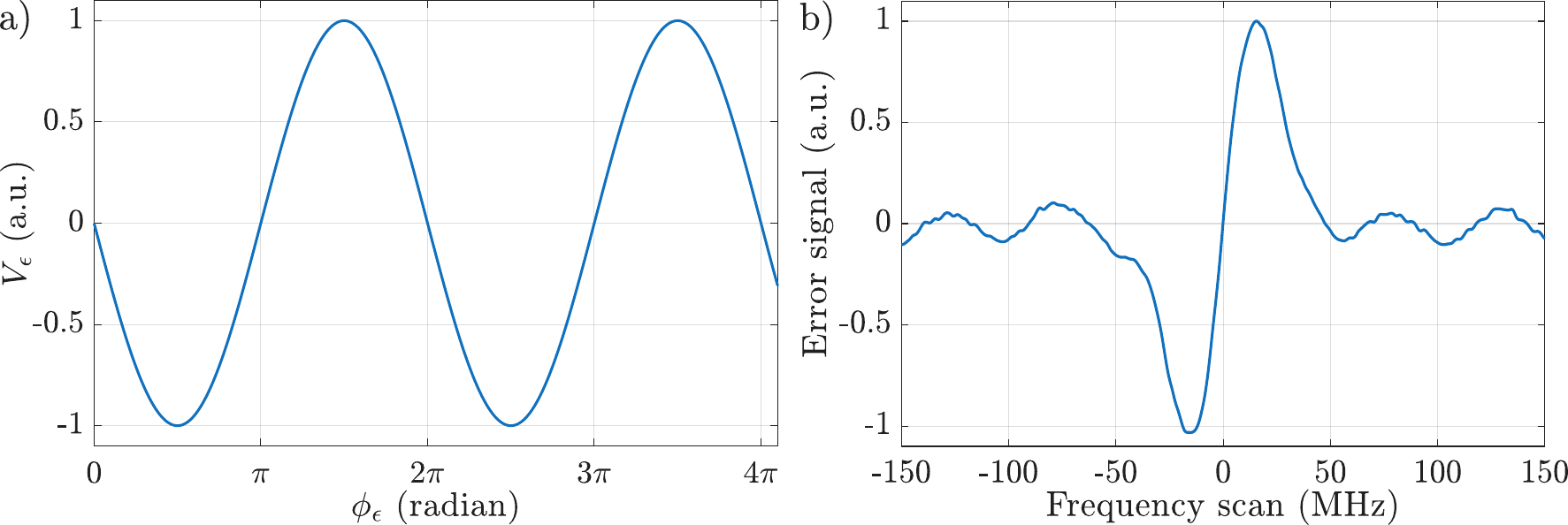}
\caption{Parasitic interference effect in error signal. a) Error signal response to the phase of parasitic beam ($\phi_\epsilon$). b) Measured error signal of the system with parasitic interference effect.}\label{fig: ER_with_PIFO}
\end{figure}

The error signal response to the phase of the parasitic beam is plotted in Fig.~\ref{fig: ER_with_PIFO}(a). The amplitude of the interference depends on the amplitude of the parasitic beam itself and how well it interferes with the main beam, i.e., polarization states and spatial modes overlap. To mitigate the effect of parasitic interference, one can reduce the amplitude of the unwanted beam. This is simple to do in a free-space system since the polarizations of the pump and probe beams are well-defined and easy to filter via polarizing optics. However, in a fiber-based system, the polarization of the beams after the fiber piece is largely dependent on temperature and stress experienced by the fiber. Polarization-maintaining (PM) PMC~\cite{Fini2014, Vidal:25} can provide better suppression of parasitic interference. However, the manufacturing process of PM-PMC is more complicated and not yet commercially available. Another way of suppressing parasitic beams is to stabilize the phase of the parasitic beams with respect to the main beam. As shown in Fig.~\ref{fig: ER_with_PIFO}(a), if the phase of the parasitic beam is stabilized close to zero crossing points, it is insensitive to the amplitude fluctuation of the parasitic beam. Figure~\ref{fig: ER_with_PIFO}(b) shows the typical measurement of the error signal with the presence of parasitic interference. The phase change of a parasitic interference ($\Delta\phi_\epsilon$) is related to the optical path length difference (OPD) between the parasitic and the probe beams, i.e., 
\begin{equation} \label{eq5}
    \Delta\phi_\epsilon = \Delta k\cdot OPD = \frac{2\pi \Delta v}{c}OPD,
\end{equation}
where $c$ is the speed of light, $\Delta k$ is the angular wavenumber difference, and $\Delta v$ is the laser frequency difference. Equation~\ref{eq5} allows us to identify the source of the parasitic beam. For example, in Fig.~\ref{fig: ER_with_PIFO}(b), the period of the interference fringe is 60\,MHz in terms of the laser's frequency scan, resulting in an OPD of 5\,m (= $c/\Delta\nu$), which is precisely the length of the PMC. This indicates that the observed parasitic interference fringe may come from the interference between the probe beam and the pump beam back reflection at the pump-PMC interface, which will be discussed later.

\begin{figure}[htbp]
\centering\includegraphics[width=9cm]{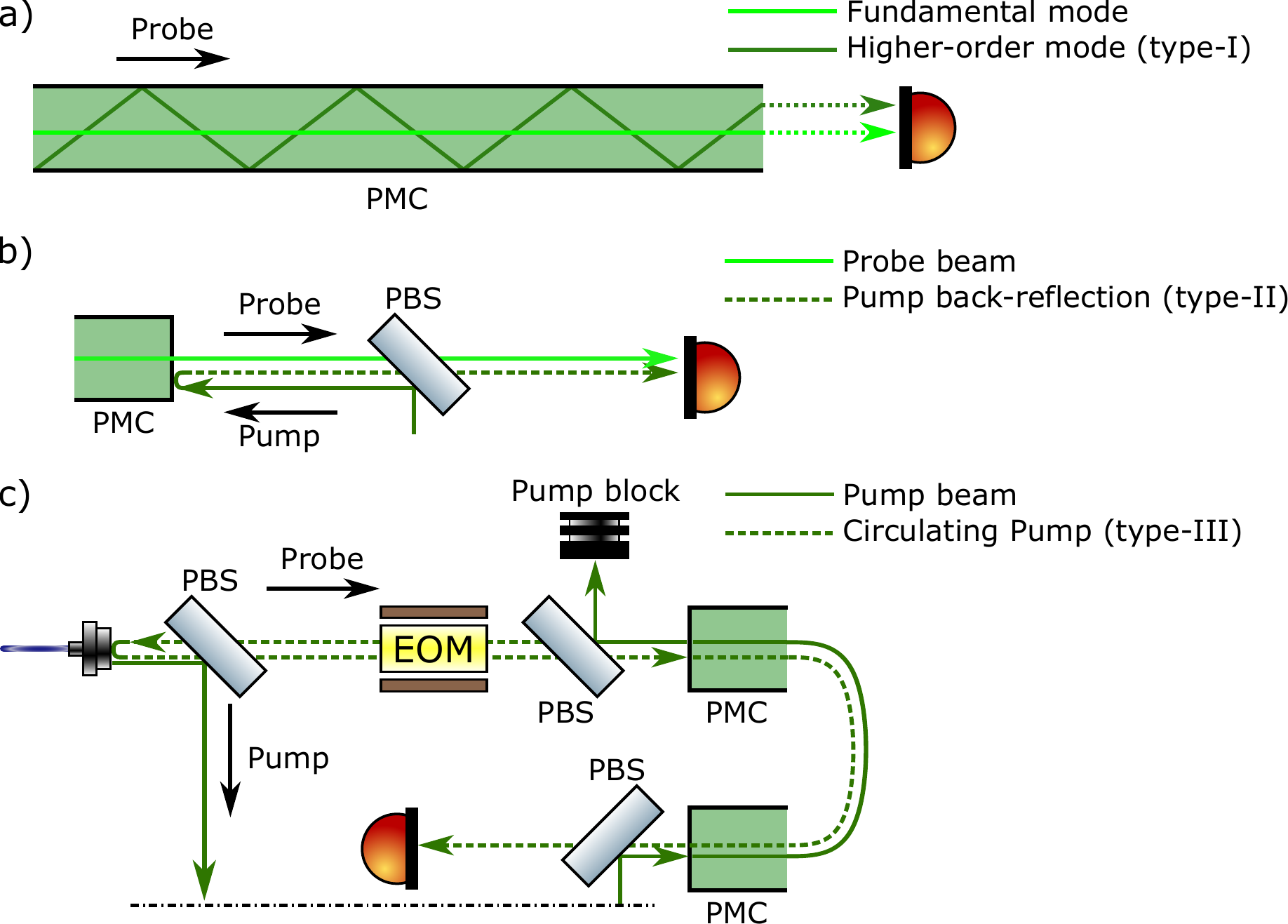}
\caption{Illustration of examples for three types of parasitic beams, some components are omitted for simplicity. a) Higher-order mode of the PMC. b) Pump beam back reflection at the pump-PMC interface. c) Circulating pump beam (dash-dotted line hides the additional pump path).}\label{fig: para_ill}
\end{figure}

In this work, parasitic interference effects are categorized into three types based on their origin. The first type of parasitic interference is between the main probe beam and ghost beams originating from the probe beams, which exhibit different phases along the optical path. The second type of parasitic interference is between the probe beam and ghost beams originating from the pump beam. Finally, the third type of parasitic interference originates from beams circulating in the optical path, where any circulating beam exhibits modulations from both the AOM and EOM. Examples of each type of parasitic beams are illustrated in Fig.~\ref{fig: para_ill}.

\subsection{Type-I parasitic interference}
The main source of type-I parasitic interference is higher-order modes coupling, as shown in Fig.~\ref{fig: para_ill}(a), which exhibits additional phase along the optical path. Any mode-matching error or alignment drift increases the coupling to the HOMs. Optimized mode-matching of the input beam to the hollow-core fiber mode is applied to minimize this effect. However, due to the collapsed region towards one end of the PMC~\cite{Light:15}, the mode pattern is not symmetric. For an input beam with a Gaussian profile, it is impossible to obtain perfect mode-matching. Thus, the coupling into HOMs is inevitable. Also, the collapsing region acts like a lens, such that the numerical aperture is larger compared to the other end of the PMC. The corresponding mode-field diameter is smaller, which makes the mode-matching of this end more sensitive to alignment. 
\begin{figure}[htbp]
\centering\includegraphics[width=0.75\linewidth]{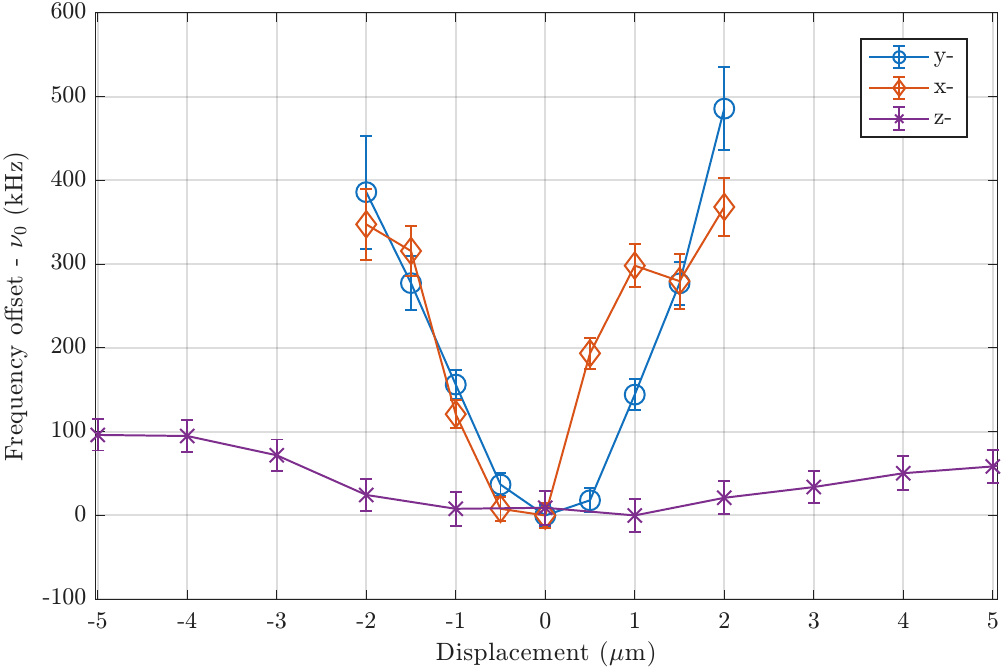}
\caption{Frequency offset of the locked laser versus position of the PMC tip with the collapsing region. $\nu_0$ is 40\,MHz away from the a1 components of R(56)32-0 of iodine.}\label{para1}
\end{figure}

To characterize how alignment affects the frequency stability, the absolute frequency of the locked iodine-filled PMC system versus the position of the PMC tip in the probe side was measured (see Fig.~\ref{para1}). The asymmetrical shape of the x-curve is due to the defects of the PMC tip, where the coupling into HOMs does not always increase as the fiber tip moves along the x direction. Only the $\rm \pm2\,\mu m$ region in the x-y plane is measured. Further away from the center, not only does the change of power ratio between the core mode and HOMs affect the parasitic interference, but also the overall power that couples into the PMC is greatly reduced, leading to a large reduction of the signal-to-noise ratio. 

As shown in Fig.~\ref{para1}, a parabolic response of the frequency offset to PMC tip position is observed. At the optimal position, minimal coupling into the HOMs gives the least frequency offset, and the system is insensitive to fiber misalignment. As the fiber tip moves away from the optimal position, the power of the HOMs increases and causes a frequency offset. This indicates that minor fluctuations in alignment near the optimal position generate little noise, while deviations away from the optimal position results in significant frequency offset and strengthens environmental couplings.

To minimize the type-I parasitic interference, a simple plastic cover over the system has proven to be effective. By covering up the system, the alignment is less sensitive to air flow and temperature fluctuations, thus reducing the coupling into the HOMs. Further suppression can be achieved by better PMC end-caps that relax the alignment requirement. Other methods, such as splicing a single-mode fiber (SMF) to the PMC\cite{Wheeler:10} to filter out the HOMs, could also reduce the fringe amplitude of the type-I parasitic interference. The splicing fixes the PMC tip position with respect to the input beam such that the system will no longer be sensitive to alignment. However, due to its small core size for 532\,nm, PMC-SMF splicing induces coupling loss, which remains a greater challenge compared to larger wavelengths. 

\subsection{Type-II parasitic interference} \label{2nd para}
\begin{figure}[htbp]
\centering\includegraphics[width=9cm]{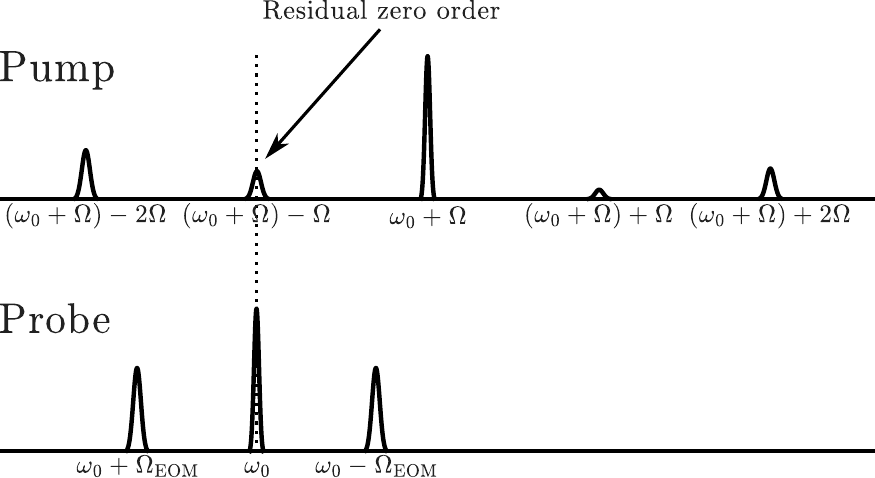}
\caption{Frequency components of the pump beam and the probe beam. The residual zero order of the AOM has the same frequency as the probe carrier. $\omega_0$: nominal frequency before the AOM. $\Omega$: AOM driving frequency. $\Omega_{\rm EOM}$: phase modulation frequency of the EOM.}\label{fig: AOM_RAM}
\end{figure}

The second type of parasitic interference is between the probe and pump beams. Ideally, these beams are separated in frequency by AOM3 as shown in Fig.~\ref{fig: locking setup}. However, residual zero order of the AOM has the same frequency as the carrier of the probe beam~\cite{Manamanni:22} such that parasitic interference occurs. The residual orders of the AOM are measured by beating the $+1$ order beam with another independent beam. A circular iris filters out all the other orders of the AOM output. The resulting frequency components of the pump beams are compared with the probe beam as shown in Fig.~\ref{fig: AOM_RAM}, where multiple residual orders are observed. 

The mode mismatch between the pump beam and the PMC results in a back-reflection at the pump-PMC interface that travels back to the PD as illustrated in Fig.~\ref{fig: para_ill}(b). Due to the residual zero order of the AOM, the back-reflected pump beam includes a frequency component at $\omega_0$, and parasitic interference occurs between this component and the probe beam. To minimize type-II parasitic interference, one can reduce the amplitude of the parasitic beam by manipulating the polarization state of the back-reflected pump beam with a quarter-wave plate. However, back-reflection occurs not only at the pump-PMC interface but also at the PMC-probe interface, which experiences the polarization drift along the PMC. Thus, it is impossible to fully remove the back-reflected pump beam by polarization manipulation. Additionally, the back-reflected pump beam goes into the input path and causes the type-III parasitic interference, which will be discussed later. 

From Eq.~(\ref{eq4}), parasitic interference depends on the phase and amplitude of the parasitic beam, where the phase is directly related to the OPD between the probe and the pump beams. Here, we introduce a phase lock loop (PLL) to minimize type-II parasitic interference by stabilizing the OPD. The back-reflected pump beam interferes with the probe at the PD at frequency $\Omega$. Since the residual zero order of the AOM follows the same optical path as the pump beam, demodulation of the signal from PD1 at $\Omega$ via a lock-in amplifier gives the OPD between the residual zero order of the AOM and the probe beam. The OPD is then stabilized by feedback control to the piezo actuator that is attached to a mirror in the pump path~(labeled as PZT in Fig.~\ref{fig: locking setup}). A polarizing beam splitter (PBS) and a quarter-wave plate (QWP) are applied to realize normal incident to the piezo mirror to minimize the misalignment from the mirror displacement. 

To estimate the remaining noise contribution of the type-II parasitic beam, we modulated the OPD between the pump and probe beams by actuating the piezo mirror. The maximum frequency change caused by the type-II parasitic interference was measured to be 200\,kHz. An out-of-loop PD was added to measure the remaining OPD with the PLL applied. A $0.7
\,{\rm mrad}/\sqrt{\rm Hz}$ out-of-loop phase stability of the PLL was measured down to 1\,Hz. Equation~(\ref{eq4}) can be simplified to a simple sine function, and applying the discriminator to both sides yields
\begin{equation} \label{eq6}
    \delta \nu_\epsilon = -\frac{A_\nu}{2}\sin(\phi_\epsilon)\simeq -\frac{A_\nu}{2}\phi_\epsilon
\end{equation}
where $\delta \nu_\epsilon$ is the frequency fluctuation caused by parasitic interference, $A_\nu$ is the maximum frequency change measured ($\simeq$ 200\,kHz), and $\phi_\epsilon$ is the measured out-of-loop phase of the parasitic beam ($\simeq0.7\,{\rm mrad}/\sqrt{\rm Hz}$). The residual frequency noise contribution from type-II parasitic interference can be estimated from Eq.~(\ref{eq6}), which yields less than $\rm 70\,Hz/\sqrt{Hz}$ above 1\,Hz. However, the HOMs also interfere with the pump beam at $\Omega$. As the phase and amplitude of the HOMs vary, residual type-II parasitic interference increases and contributes to frequency instability for frequencies below a few millihertz. 
\begin{figure}[htbp]
\centering\includegraphics[width=0.75\linewidth]{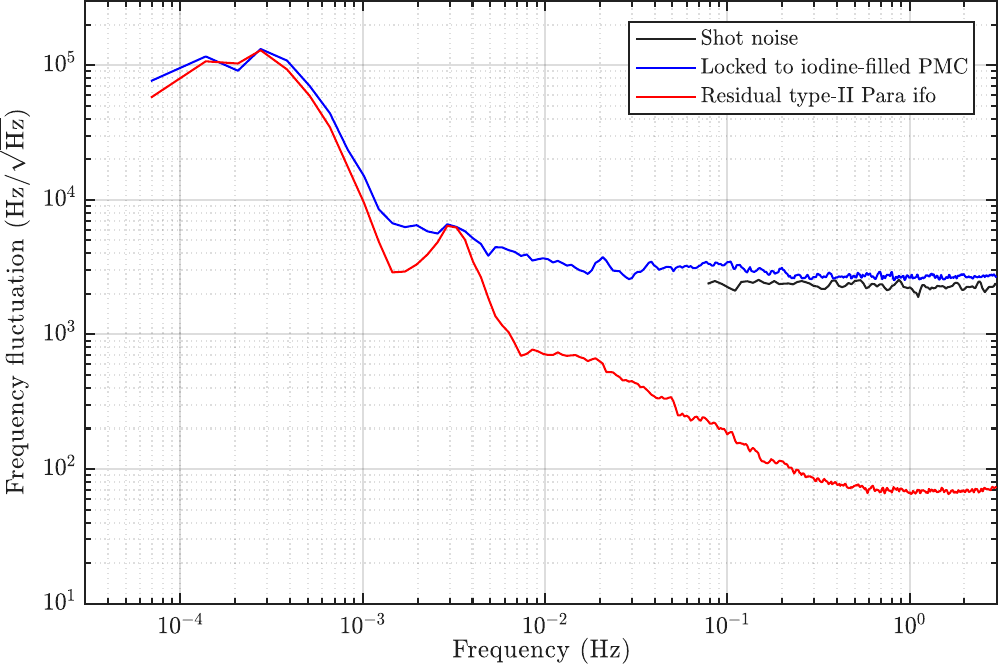}
\caption{Frequency stability of an initial test. Frequency noise contributions from residual type-II parasitic interference and shot noise are plotted.}\label{fig: re_2nd}
\end{figure}

Figure~\ref{fig: re_2nd} shows the measurement result of an initial test. The out-of-loop result of the PLL phase is measured and converted to residual type-II frequency noise by applying Eq.~(\ref{eq5}) as described above. The system is limited by residual type-II parasitic interference for frequencies below 1\,mHz. Between 1\,mHz and 100\,mHz, other types of parasitic interference contribute to frequency noise. Since an optical splitter is added for out-of-loop phase measurements, the optical power available for frequency stabilization is reduced. Consequently, the shot noise contribution is higher, and limits the measurement in the frequency band above 100\,mHz. For our final measurement in Sec.~\ref{result}, the splitter was removed to obtain the best noise floor of the system by using the maximum amount of power available. 
Better end-capping of the PMC can result in less back reflection, which reduce the type-II parasitic interference. Polarization-maintaining PMC is also a potential solution, where polarization states are better controlled and, consequently, the use of polarization optics is more effective.

\subsection{Type-III parasitic interference}
Type-III parasitic interference is caused by beams that circulate in the optical path. The polarization state of the PMC output is sensitive to temperature and vibration, making it difficult to separate and dump the beams using PBS. Back-propagating beams reflected at any optical surface, especially at the fiber interfaces, will circulate in the optical path. An example of a circulating pump beam is shown in Fig.~\ref{fig: para_ill}(c), which is phase-modulated and carries the dispersion information of the iodine. A similar back reflection happens to the probe beam as well. Multiple orders of phase-modulated beams and frequency-shifted beams are able to reach the PD and interfere at the demodulation frequency. Consequently, interference fringes appear in the error signal. Since the circulating beams share common optical paths, this effect can not be removed by the phase lock loop described in Sec.~\ref{2nd para}. 
\begin{figure}[htbp]
\centering\includegraphics[width=0.75\linewidth]{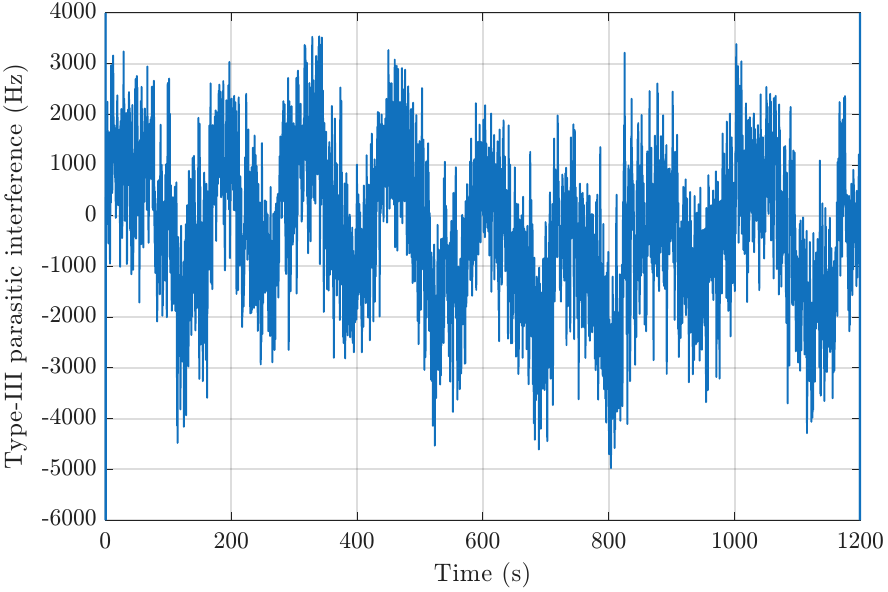}
\caption{Time series of frequency measurement with the type-III parasitic interference effect. Frequency data is plotted with respect to the beat note frequency at the beginning of the measurement. The amplitude of the fringes is 2.7 kHz.}\label{fig: 3rd ifo}
\end{figure}

To characterize the contribution of the type-III parasitic interference, we measured the power of the circulating beams. After inserting isolators in either the pump or probe path, the frequency fluctuations of the third-type parasitic interference were measured by recording the beat note frequency of the locked PMC system compared to the commercial free-space system. Note that the PLL was engaged to remove type-II parasitic interferences. Figure~\ref{fig: 3rd ifo} shows the time series of the measured frequency with type-III parasitic interference from the pump beam. A similar measurement was done with an isolator on the probe side. According to Eq.~(\ref{eq4}), fringe amplitude scales linearly with the power of the circulating beam. The power-to-fringe amplitude factors are then calculated to be $\rm200\,Hz/\mu W$ and $\rm100\,Hz/\mu W$ for the pump and probe beams, respectively.

To minimize the type-III parasitic interference, two Faraday isolators with 41\,dB isolation were inserted in the setup as shown in Fig.~\ref{fig: locking setup}. The power levels of the circulating beams were reduced from tens of $\mu$W to a few nW in both the probe and pump beams. Applying the power-to-frequency coupling factors, the remaining frequency fluctuation amplitude is less than a few Hertz. Since the phase change of the circulating beams mainly comes from the PMC, the PLL feedback signal described in Sec.~\ref{2nd para} can be used to estimate the phase fluctuation of type-III parasitic beams. The remaining noise contribution of the type-III parasitic interference can be estimated by applying Eq.~(\ref{eq6}), resulting in a noise level that is two orders of magnitude lower than the shot noise. 

\section{Frequency stability results} \label{result}
\begin{figure}[htbp]
\centering\includegraphics[width=0.75\linewidth]{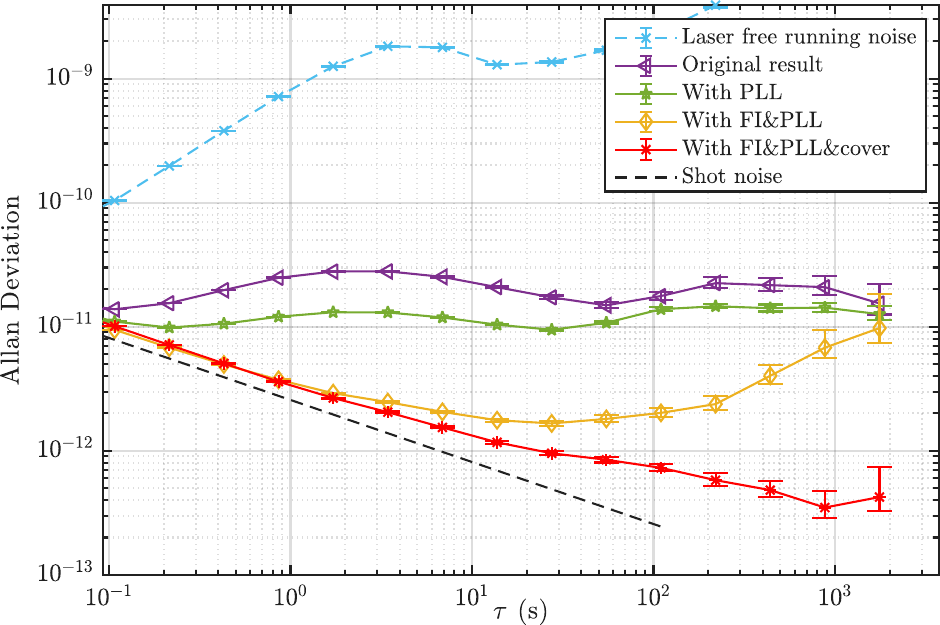}
\caption{Allan deviation curves of the iodine-filled PMC system. The parasitic interference effects are mitigated by applying isolators, a phase-lock loop, and a plastic cover. Step-by-step improvements are shown as labeled. FI: Faraday isolators. PLL: phase lock loop.}\label{fig: AL}
\end{figure}

The noise suppression methods described in Sec.~\ref{noise} are applied to the system. Step-by-step improvements of the frequency stability of the locked system in terms of Allan deviation are shown in Fig.~\ref{fig: AL}. The free-running noise of the laser is shown as a reference. The original result shows a frequency stability of $10^{-11}$. The type-II parasitic interference effect was mitigated by applying the phase lock loop, resulting in a factor of two improvement. After applying the Faraday isolators, type-III parasitic interference is minimized, and an order of magnitude stability improvement was seen within 1\,s to 1000\,s integration time. The type-I parasitic interference effect remains the main noise source for integration times longer than a few seconds. After applying a cover around the system, air flow and temperature-induced alignment drifts and OPL changes are reduced, which mitigates type-I parasitic interference and greatly improves the long-term frequency stability. The system is shot noise limited for integration times below 10\,s, where the remaining type-I parasitic interference causes long-term frequency drifts. This is due to alignment shifts and temperature fluctuations that cause power and phase change of the higher-order modes. Nevertheless, stability levels below $10^{-12}$ for integration times between 10\,s and 1000\,s are achieved for the first time in gas-filled PMCs. The fractional frequency stability of the system is as low as $3.5\times10^{-13}$ at 1000\,s integration time.

\section{Conclusion} \label{conclusion}

We have demonstrated fractional frequency stability of $3.5\times10^{-13}$ for integration times around 1000\,s, which, to our knowledge, is the best frequency stability reported using an iodine-filled PMC fiber frequency reference system. This performance paves the way for future all-fiber-based frequency reference systems, which are particularly attractive for space applications where size and weight are critical constraints.

We have identified three types of parasitic interferences that limit frequency stability. To address these, we designed and implemented suppression methods based on parasitic amplitude reduction and phase control, achieving more than an order of magnitude improvement for integration times longer than a few seconds. For gas-filled PMCs, parasitic beams cannot be fully eliminated due to the inevitable coupling of light into higher-order modes. Therefore, phase stabilization of the parasitic beams provides a feasible approach to mitigate their impact on the system. When combined with parasitic amplitude reduction techniques, these methods can greatly improve frequency stability. Nevertheless, type-I interference is still the dominating noise source for time scales longer than 10\,s.

Several mitigation strategies can further enhance performance at long time scales. Type-I parasitic beams can be reduced through temperature stabilization of the PMC fiber. The temperature set point can be chosen to minimize the phase difference between the main mode and higher-order fiber modes, thereby improving system frequency stability. Alternatively, type-I effects can be minimized by implementing active feedback to maintain PMC alignment and thus minimize coupling into higher-order modes. Finally, improved PMC fiber ends will not only reduce type-I parasitic interference but also increase coupling efficiency into the fiber, resulting in a larger discriminator and a reduced shot-noise limit.

\begin{backmatter}
\bmsection{Funding}
The authors acknowledge financial support from the National Aeronautics and Space Administration (NASA) through grant NASA IIP: 80NSSC24K1097; and Department of Defense (DOD) through grant W911NF2410383.

\bmsection{Acknowledgment}
The authors would like to thank GLOphotonics for fruitful discussions and PMC modifications; and Cl\'ement Go\"icoech\'ea
for providing insights on hollow-core fibers' mode-matching.

\bmsection{Disclosures}
The authors declare no conflicts of interest.

\bmsection{Data Availability Statement}
Data underlying the results presented in this paper are not publicly available at this time but may
be obtained from the authors upon reasonable request.
\end{backmatter}

\end{document}